# Bloch Surface Wave-atom Coupling in Periodic Photonic Structure

M. ASADOLAH SALMANPOUR, M. MOSLEH, S. M. HAMIDI*
*Magneto-plasmonic Lab, Laser and Plasma Research Institute, Shahid Beheshti University, Tehran, Iran*
*: m_hamidi@sbu.ac.ir

**Abstract:**

Considering efforts for hot atom vapor-nanophotonic integration as a new paradigm in quantum optics, in this paper, we introduce 1D photonic crystal-Rb vapor cell as structure with miniaturized interaction volume. The Bloch surface wave (BSW) excited on surface of a photonic crystal as electromagnetic hosting photonic mode, and altered the optical response of Rb atoms in the vicinity of surface. Coupling of atomic states with BSW confined modes would lead to quantum interference effects and results in nonlinearities in resonant coupling of atoms with BSW. We show Bloch surface wave induced transparency is highly stable under a change of incidence angle. Our results show slight changes in transitions detuning's due to nonlinear interactions like the Casimire-Polder effect under change of localized density of optical states.

## I. Introduction

Interests on design and construction miniaturized new-born quantum devices and atomic chips motivated scientists to investigate atom-light interactions in nanostructures. Miniaturization and integration of nanophotonic-atomic physics would result in enhancement of interactions of atoms with confined light field as a consequence of scale down of the efficient interaction volume. Due to the nonlinear phenomena that may occur, studying the interaction of atoms and light in the vicinity of high-intensity and limited fields in a smaller volume becomes more important. Promising paradigms For miniaturization based on integration of atom-nanophotonics context exploits different types of evanescent fields, regarding favourable properties. The interaction of the atom and tail of the evanescent field has been investigated in structures such as integrated hollow waveguides [1,2], solid core waveguides [3, 4], and integrated ring resonators [5,6] have been designed. For example one of the extend studied types of the evanescent fields is surface plasmon polaritons (SPP) which confines electromagnetic field in the planar interface between a thin metal film and a desired dielectric medium or our aim atomic media[7]. One of the features of SPP's is highly modulable resonant fields depending on excitation configuration that would make it candidate in specific atom-SPP interaction like all-opticl switching [8], and our recently work as multifunctional logic gates [9]. SPP's mode dissipation by temperature increasion caused from ohmic losses of metals[10], introduce new bottlenecks for many atomic hot vapor applications based on SPP' evanescent field spectroscopy. Here we want to introduce Bloch surface waves (BSW) as a candidate evanescent field to be coupled to hot atomic vapor, considering neglible changes under changes of excitation configuration, specially incidence angle and temperature changes.

BSW$_s$ take advantage of inconsiderable ohmic loss with possibility to propagte at the interface between periodic dielectric multilayer and a surrounding medium in the scale of millimeter length. BSW's also, represent chance to generate great enhancements of TE or TM polarized evanescent fields [11,12].

BSW$_s$ in photonic crystals (PCs) are known as very broad applications in sensing [13-16], Control of emission properties of emitters [17], surface-enhanced Raman scattering [18,19], enhanced fluorescence detection[20,21], trapping Au particles[22] and BSW-based integrated photonic platforms, such as 2D lens[23], optical resonator[24,25], flat reflector[26], nanocoupler[27] and so on. As a consequence, by designing nanostructure on the surface of the photonic crystal, Bloch surface modes can be guided in polymeric ridge [28], curved waveguides [29], and Optical resonators [25]. Feasibility of stimulation of the BSW's in resonator configuration promises facilities for investigation of the interactions in atom-cavity coupled systems, without dealing with complexities like couplers for excitation of resonator modes.

Numerous publications show some fundamental phenomena arising from coupling of hot or cold atoms to confined field [30,31]. Comparing complexities of atom trapping configurations with compatibility of hot vapor systems to be integrated in guided light systems, inspires many studies onto investigation of hot atomic vapors hybridization with confined light. Coupling of atomic states with photonic confined modes would lead to quantum interference effects and results in phenomena like as Fano resonance, EIT-like resonance, and nonlinear interactions[8,32,33]. Resonant coupling of hot flying atoms with the photonic modes in solid substrates depends on properties of both sides of coupled structure. Optical response of hot atom-confined mode hybridization will undergo variations by shifts on interaction configuration like, interaction time of atoms with photonic field, or photonic mode properties specially, field strengths and resonance frequency.

1D PC-atomic cell structure and BSW-atomic state hybridization introduced in this report is introduced for first time, to best of our knowledge. we consider the Bloch surface wave as a candidate for the miniaturization of atomic vapor cells, and try to investigate the effects of pivotal properties of BSW excitation configuration on coupling of the BSW-atom and resultant linear and nonlinear interactions between them. We believe higher values of efficiency in the coupling of Bloch surface waves and atoms can be achieved in the newborn BSW-atomic cells.

## II. Material and Method:

At first glance, we fabricate 1D periodic photonic crystal structure by BK7/$(SiO_2/ZrO_2)^{12}$/$SiO_2$ construction. The design wavelength in the fabrication process was set to 808nm in quarter wavelength selection to reach the D2 line of Rb vapor in the PBG region.

Finally, the transmission and reflection spectrum of the fabricated sample was taken and simulated with the 2*2-based transfer matrix (TMM) method. Considering the Bloch theorem, a wave propagating through a periodic medium may exit only if it keeps the same amplitude but with a constant phase shift across the photonic band gap (PBG).

BSW-atomic cell was fabricated with the aid of a rubidium-filled glass cell and was attached to the photonic crystal by epoxy in such a way that the photonic crystal surface is in the vicinity of rubidium vapor. The substrate of PC was contacted with the prism using an index match (Figure(3a)). To reach proper density of Rb atomic vapor, the fabricated cell was heated up to $70^0$c by our homemade heater.

The wave vector of BSW lies beyond the light line, and it can't couple in the air. Thus, TIR configuration is needed for BSW excitation [11]. Furthermore, to excite more efficient surface Bloch wave in the PBG of structure, we use kerstchmann configuration to cover our purpose incidence angles. To evaluate the BSW wave in the sample, the reflection spectrum of the sample was measured as a function of the incident angle of light, at 795 nm, on the reflecting face of the prism. After that, the sample was put into the frequency modulation spectroscopy setup as shown schematically in Fig.1. In this experimental setup, frequency-modulated light from distributed feedback (DFB) diode laser , with wavelength of 795 nm Corresponding to the D1 line of Rubidium, after a pass from Glan Taylor prism, reflects from prism coupled BSW-atomic cell and reaches to the lock-in amplifier (LIA) connected detector. Then lock-in amplifier demodulated and amplified this signal which was locked on the frequency of the laser driving current generated by a function generator. In the measurement process, the laser's optical frequency was scanned across the hyperfine structure of Rb. (As shown schematically in Fig. (1)).

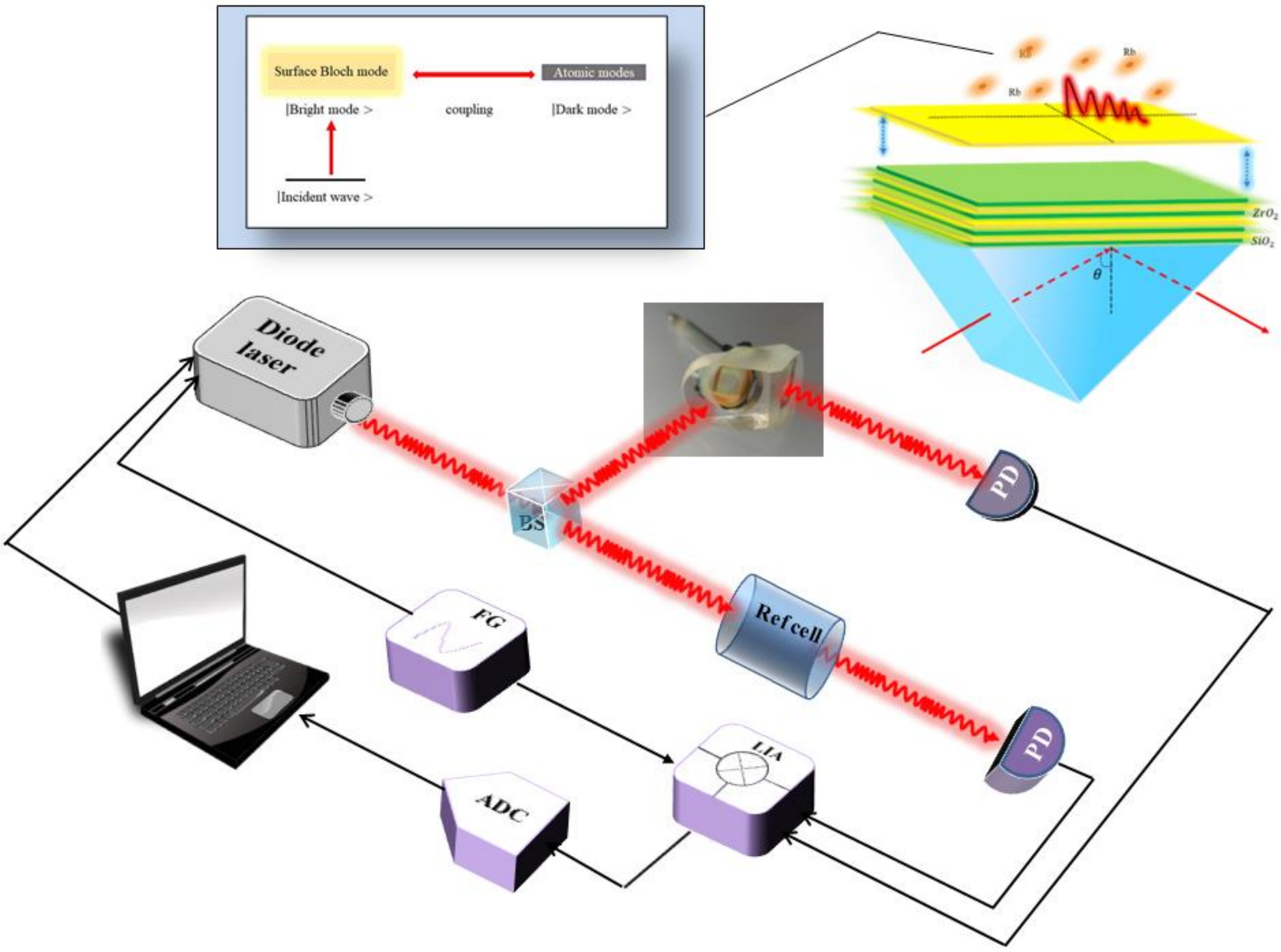

*Figure1: Schematic diagram of experimental setup used for measure reflectance from boundary of coated PC and Rb vapor. Inset reperesents level scheme schematics for EIT-like in a three-level system. The incedent mode coupled directly to the bloch surface mode which called bright mode. coupling between incident mode to the atomic states is impossible thus atomic states are dark mode. The coupling between the Bloch surface state and the discrete atomic state is also possible with a coupling coefficient.The two possible trajectories, namely,* $|incident\ mode\rangle \rightarrow |bright\ mode\rangle$ *and* $|incident\ mode\rangle \rightarrow |bright\ mode\rangle \rightarrow |dark\ mode\rangle \rightarrow |bright\ mode\rangle$ *interfere destructively and lead to the EIT-like phenomena.*

In frequency modulation measurements, the output signal from the lock-in amplifier is proportional to the derivative of the atomic absorption signal. Therefore, the measured spectrum from a Lorentzian atomic absorption is dispersion-like. Because, in many cases, people use this dispersion, like the signal from LIA, as an error signal in control systems. In this paper, we call 1$^{st}$ harmonic output of our FM spectroscopy measurements “error signal”.

## III. Results and Discussion:

The reflective index distribution and transmittance spectrum of our PC Simulation and measurement results are shown in Fig (2a,b) Respectively. This confirms the PBG in the range of 700 to 900nm. This PBG covers our aim BSW in the D1 line of Rb vapor. The slight difference between the experimental and simulation results is due to the disorder in the thickness of the layers during the deposition process.

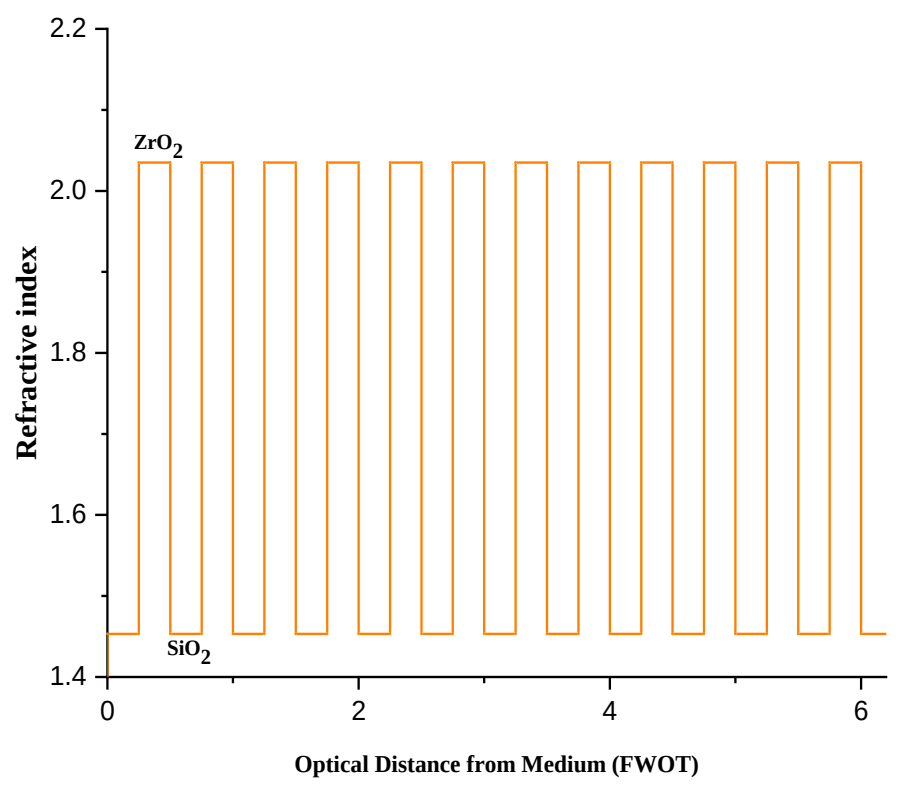

(a)

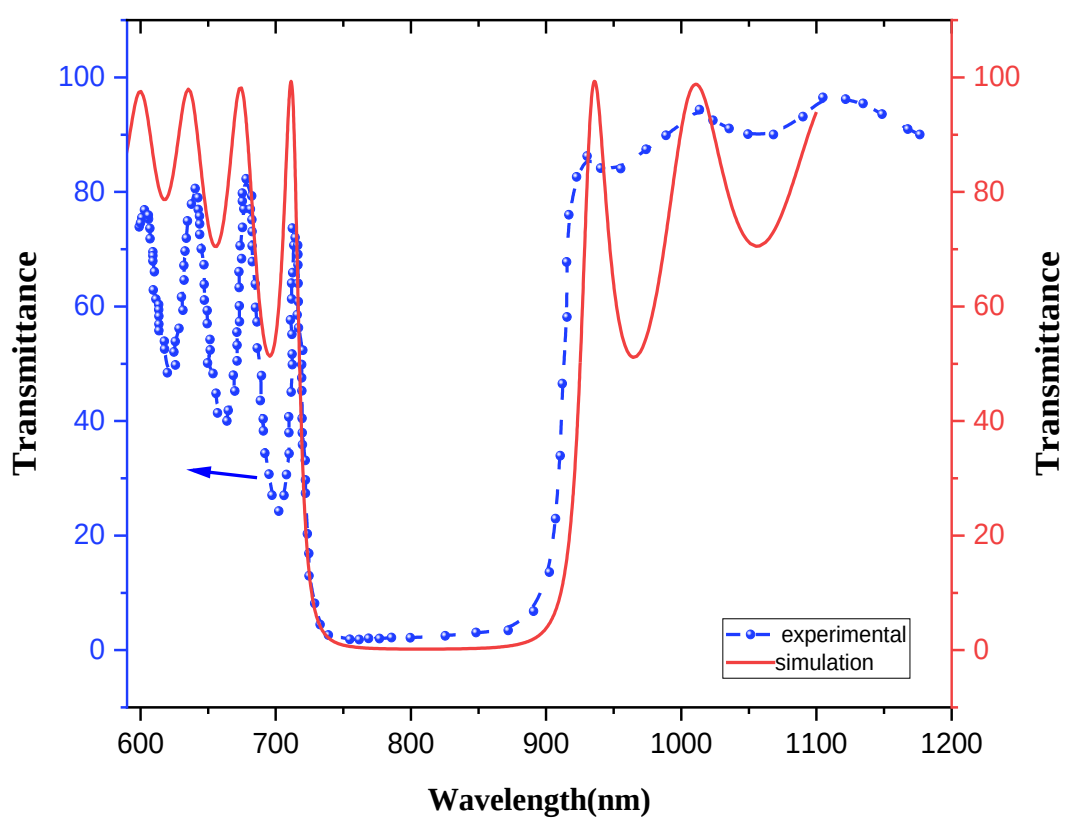

(b)

*Figure(2): (a) reflective index distribution of photonic crystal. (b) Transmission spectrum as function of the vawelength, simulation( red line) and experimental(dashed blue line).*

After evaluating the main PC, we record the BSW in the PBG by angular modulation in the Krestchmann configuration, as shown in Figure (3b). This sample's reflection spectrum was measured without considering the effect of rubidium atoms at the angles after the total internal reflection angle (42°). The BSW-atomic cell was mounted on a rotation stage, and using the wavelength of 795, which corresponds to the Rb D1 line, the reflection spectrum of the sample was measured. As shown in figure (3, b), three reflective dips are observed, two related to guide modes in crystal photonics and one related to Bloch surface mode. The narrow and deep resonance at an angle of 64.50° indicates a BSW mode whose field distribution is maximized in the last layer of the PC, and light confinement occurs on the side of the substrate PC. The other dip is related to the excitation of the Bloch-guided modes associated with the band edge photonic crystal. Guide modes are not surface modes because their field distributions are maximized in the center of the multilayer [34].

Figure (3c) shows the reflection spectrum of our BSW-atomic cell's response at a 64.50-degree Incidence angle while the laser current swept to include the transitions corresponding to $F_g = 2 \rightarrow F_e$ for $^{87,85}$Rb, $F_g = 3 \rightarrow F_e$ for $^{85}$Rb and $F_g = 1 \rightarrow F_e$ for $^{87}$Rb. A comparison of reflection from the BSW-atomic boundary (red line) with the absorption of flying bulk vapor atoms shows two error signals are symmetric regarding the intensity axis. As the output of LIA in FM spectroscopy (error signal) is derivative of real data, inversion of intensity changes in reflection compared to absorption data means BSW-atomic coupling leads to induced transparency in hyperfine line detuning. In our hybrid system, induced transparency phenomena is formed due to the coupling of a BSW (bright mode) and atomic states (dark mode) as shown schematically in the inset of Fig. 1. Coupling description using the model of two coupled classical damped harmonic oscillators is as follows [35]:

$$\begin{pmatrix} \omega_b - \omega - i\gamma_b & g \\ g & \omega_a - \omega - i\gamma_a \end{pmatrix} \begin{pmatrix} x_b \\ x_a \end{pmatrix} = i \begin{pmatrix} E_0 \\ 0 \end{pmatrix}$$

Here $x_b$ and $x_a$ are oscillator amplitudes (charge displacement) of BSW and atomic mode, $\omega_b$ and $\omega_a$ are central frequency of Bloch surface resonant and atomic resonant, $\gamma_b$ and $\gamma_a$ damping parameter of Bloch surface mode and atomic mode, ω is the Laser frequency of incident light, and g is the coupling constant between two oscillators, which determines the strength of the interaction between the Bloch mode and the atomic mode. Since the laser light has no interaction with the atomic environment, on the right side of the equation, the external force to the oscillator corresponding to the atom is zero. The number of layers of the photonic structure of the crystal does not allow the laser light to reach the atomic environment. In this hybrid system, it is much smaller than the damping parameter of Bloch and atom modes ($g \ll \gamma_b$ or $\gamma_a$). The atomic oscillator amplitude is

$$x_a(\omega) = \frac{igE_0(\omega_a - \omega - i\gamma_a)}{(\omega_b - \omega) \times (\omega_a - \omega) - i\gamma_a(\omega_b - \omega) - i\gamma_b(\omega_a - \omega) - \gamma_b\gamma_a - g^2}$$

By multiplying the charge by the displacement of the atomic charge, it is possible to calculate the Polarizability, P, which is proportional to complex susceptibility $\chi$ and external field E. Since$\gamma_b \gg \gamma_a$, $x_a \propto \chi_a$, the susceptibility of the Bloch mode is not affected by this coupling, but the susceptibility of atoms changes due to this coupling according to equation 1.

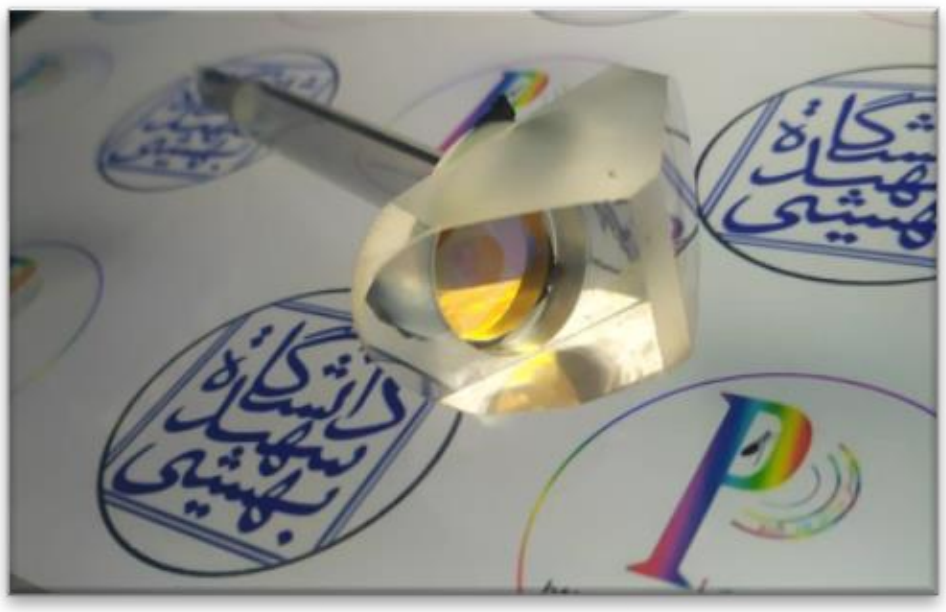

(a)

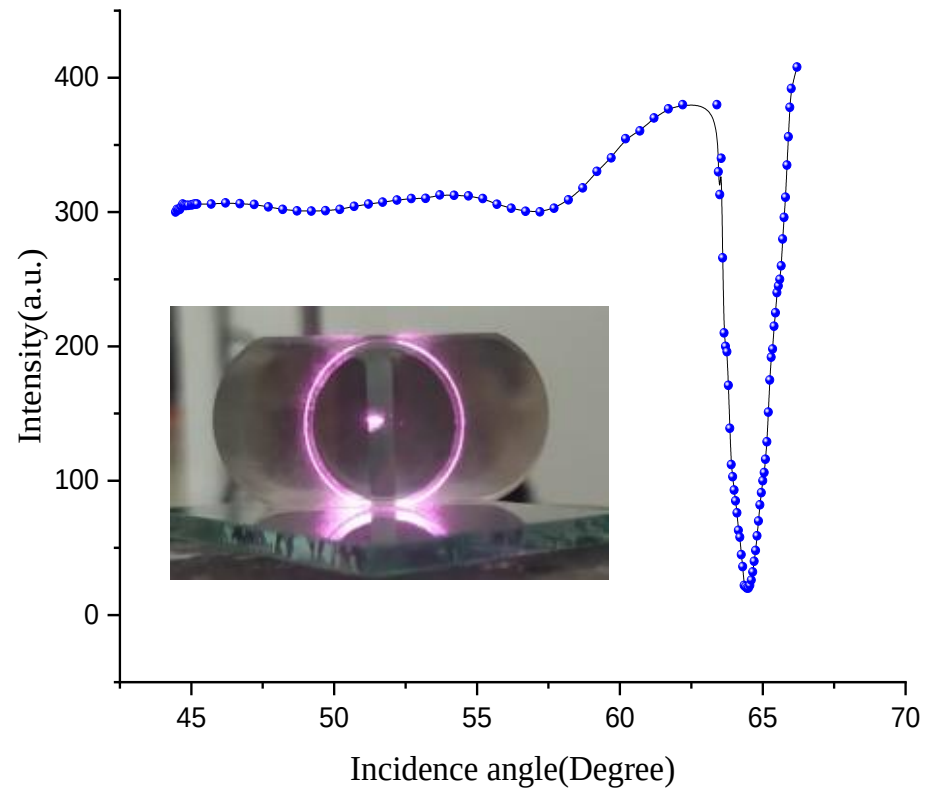

(b)

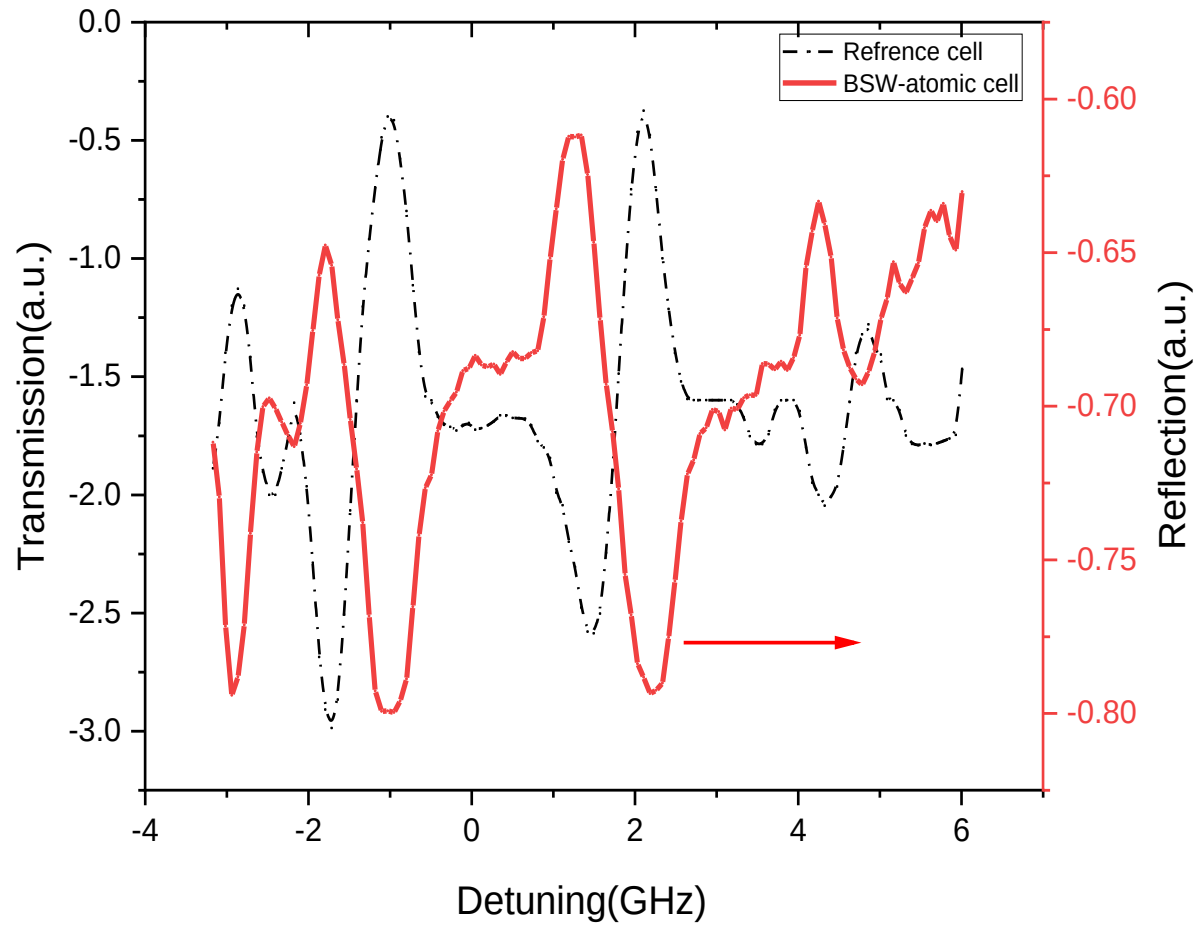

(c)

*Figure(3): Design and characterization of a BSW-atomic vapor cell. (a) Photograph of BSW-atomic rubidium vapor cell. (b) Angular reflectivity of proposed cell at t the angles after the total internal reflection angle (42°). Inset shows Photograph of the bare photonic crystal coupled prism. (c) reflection of the BSW-atomic system at Bloch surface resonance angle(red line). Transmission through a reference Rb cell is presented by the dashed black lines.*

In the deep of the Bloch surface wave(Bloch resonance angle), $\omega_b = \omega_a = \omega$, therefore, susceptibility of the atom will be proportional to $\mathcal{X}_a \propto -\frac{g\gamma_a}{\gamma_a\gamma_b - g^2}$ Using equation 1. Since the energy absorption is related to the imaginary part of the susceptibility of the atomic mode and under the mentioned conditions, we do not have an imaginary part, so the atom's absorption is zero, and the atom becomes transparent.

Due to the damping coefficient of the oscillator, even if the system is in its natural motion (without external force), the natural frequency of the oscillator is $\omega_f = \sqrt{\omega^2{}_b - \gamma^2{}_b/4}$. The dynamic change of the damping coefficient of this oscillator causes the modulation of the resonant frequency of the Bloch surface mode. We did measurements at different angles to observe the effect of a change in the angle of incidence of light in the reflection response of a BSW-atomic coupled structure (Fig 4). The results show a slight change in the angle of incidence of light about the BSW resonance angle, which has no considerable effect on reflected intensity. In fact, induced transparency phenomena are seen without perturbations in all different angles. As atom states are almost stable under changes of measurement geometry, from two coupled states analysis, we know this behavior is only valid if BSW resonance frequency doesn't change under a change in angle of incidence. The most common reason causing a change in the resonance frequency of the surface wave under a change of angle of incidence is a change in damping of resonant mode [*8*]. We believe the stability of BSW damping under a change in the angle of incidence is related to the origin of the Bloch mode.

The net damping of a surface wave (including BSW) consists of three different types, known as leakage damping ($\gamma_L$), internal damping of materials carrying surface waves ($\gamma_{int}$) and Surface scattering ($\gamma_{scat}$). Leakage damping results from energy leakage from BSW modes back into the prism and fully depends on the angle of incident light. Due to the bandgap of the multi-layered structure of the photonic crystal, a small amount of energy returns to the prism, therefore changing on an angle around the Bloch resonance angle, $\gamma_L$does not change much in energy leakage, nor does the leakage damping. On the other hand, the internal damping and surface scatterings arising from inherent losses and the ingredients of photonic crystal and interfacial roughness don't change under a change of angle of incidence. Considering these facts, a very good approximation, the total damping of BSW mode under the change of angle, would be considered zero [11].

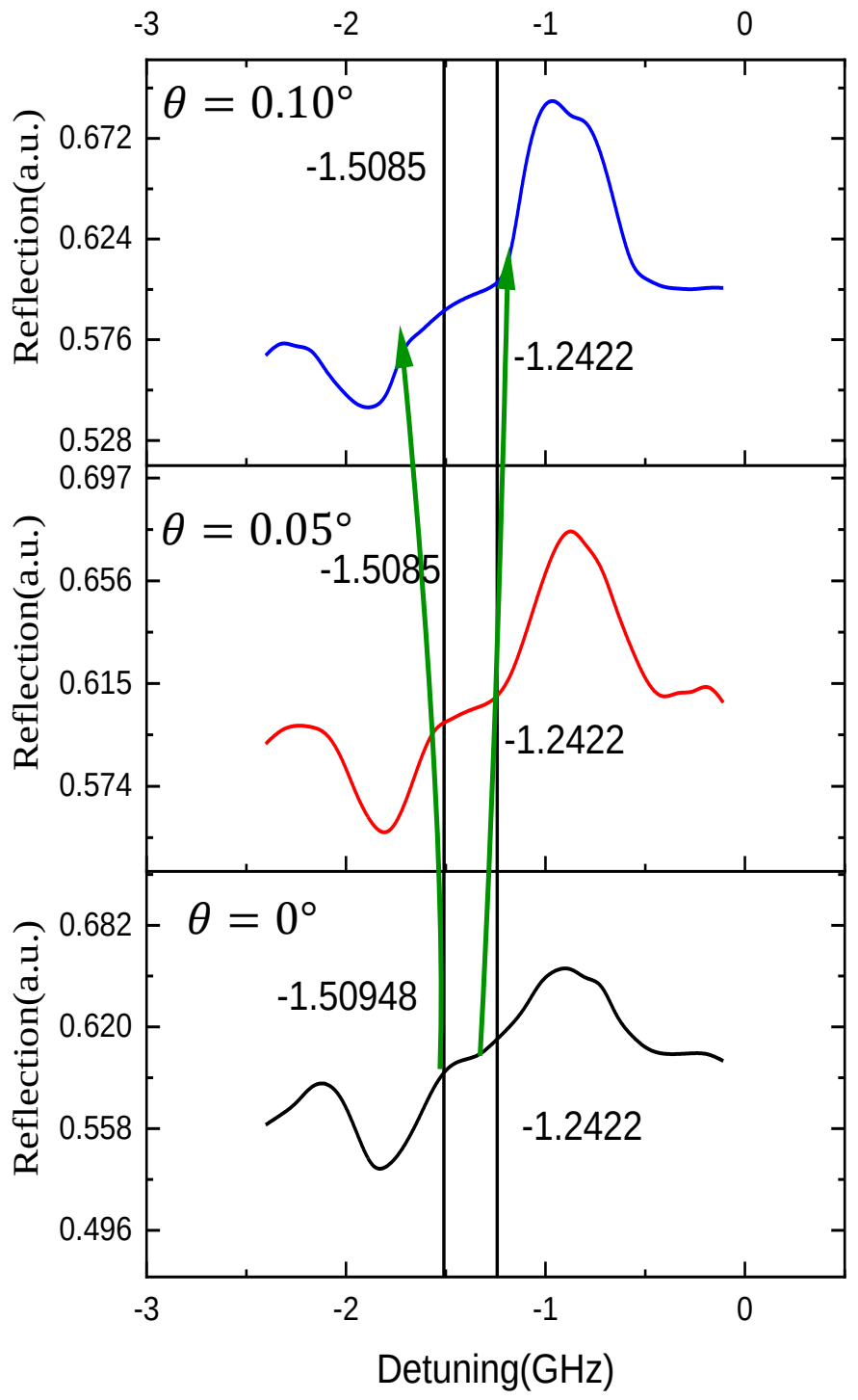

*Figure(4): Measured reflection spectrum as a function of frequency detuning at incidence of light angular modulation of selected transition line about 0.1 degree. two vertical lines represent $F_g = 3 \rightarrow F_e = 2$and $F_g = 3 \rightarrow F_e = 3$ transitions of $^{85}Rb$.*

As discussed, hot atomic vapor-nanophotonic integrated systems are candidates for scalable and easy-to-run quantum devices. One of the important considerations in integrating atom-nanophotonic is the mutual effect of optical mode and atomic states on each other, Although any change in properties of atomic states is of greater importance. In our proposed hybrid system, the linewidth of Rb atomic transitions in the reference cell is 688.2 MHz, but in the reflection spectrum from a coupled state, the linewidth is 993.6 MHz. The main reasons for these broadenings are Doppler broadening and Transient time broadening. Increased Doppler broadening caused by bigger *k* vector of evanescent wave (BSW) compared with free space light ($k_{SBW}.v > k_{light}.v$). The transition time broadening gets dominant when atom-surface wave interaction time gets shorter than the net lifetime of an atom under natural and doppler broadenings. Because interaction time is inversely proportional to the interaction length (spatial expansion of surface wave), an increase in the interaction length of the atom-surface wave leads to an increase in interaction time. As a result, the transition time broadening gets ineffective. Since the spatial distribution of the Bloch surface mode along the interface is several microns, the reduction in the transition time broadening of atoms touched by BSW is much shorter than other surface waves, such as highly damping surface plasmon polariton. We expect BSW mode to be a good candidate for the miniaturization of hot vapor atomic devices.

Photonic modes in nanophotonic structures possess features like high-intensity electromagnetic fields in small mode volumes that would enhance the interaction of atoms with the electromagnetic field. Although the large cooperativity of atoms with photonic modes is manifested in many quantum mechanical applications, the large gradient of an intense electromagnetic field in the vicinity of a surface of nanostructures would make atom-electromagnetic field interactions sophisticated. In the case of atom-BSW coupling, a modulation in the coupled state would be possible by modulation of the angle of incidence of light. In our proposed structure, a change in incidence angle will strongly affect the local density of photonic states of nanophotonic mode (LDPS). It is well known that a change in LDPS would change the electronic states of the atom [36]. Here to show the impact of LDPS on atomic transitions, we slightly swept the incidence angle of light on BSW-atomic vapor structure and did measurements in a small detuning range of $F_g = 3 \rightarrow F_e = 2$and $F_g = 3 \rightarrow F_e = 3$ transitions of $^{85}$Rb. Fig 4 shows the error signal when reflection from 1D-PC is minimum; we call it zero resonance angle. An interesting phenomenon observed here is a slight change in reflected intensity exactly in the position of $F_g = 3 \rightarrow F_e = 2$and $F_g = 3 \rightarrow F_e = 3$ transitions of $^{85}$Rb. It must be noted that Doppler and transient time broadenings are still dominant. Still, we believe this increased spectral resolution would result from the saturation of the atoms in the must probable thermal velocity in the vicinity of a highly intense evanescent field. From Fig4 (a, b & c), its clear slight change of angle of incidence from resonance angle makes a small shift (several MHz) on transitions detuning. To understand this behavior, we consider the effect of nonlinear interactions like the Casimire-Polder effect and the change of LDPS. Our results confirm that BSW's arrangement of photonic states exited in resonance angle neutralize any shift on atomic transitions.

## IV. Conclusion:

In sum, we fabricate a BSW-atom cell based on one-dimensional photonic crystal and rubidium hot vapor for the first time. Due to the BSW's low loss and high penetration depth, we assume the main casmier-polder effect onto the LDPS and slight shift and change in the transition line resolution and broadening. We believe the stability of BSW damping under a change of angle of incidence as leakage damping ($\gamma_L$), internal damping of materials carrying surface waves ($\gamma_{int}$) and Surface scattering ($\gamma_{scat}$). In addition, Doppler and transient time broadenings still are dominant. Still, we believe this increased spectral resolution would result from the saturation of the atoms in the must probable thermal velocity in the vicinity of a highly intense evanescent field.

**Coflict of interest:**

There is no any conflict of interest.

## References

[1] Schmidt H, Hawkins AR. Atomic spectroscopy and quantum optics in hollow-core waveguides. Laser & Photonics Reviews. 2010 Nov 2;4(6):720-37.

[2] Yang W, Conkey DB, Wu BI, Yin D, Hawkins AR, Schmidt H. Atomic spectroscopy on a chip. Nature Photonics. 2007 Jun;1(6):331-5.

[3] Ritter R, Gruhler N, Pernice W, Kübler H, Pfau T, Löw R. Atomic vapor spectroscopy in integrated photonic structures. Applied Physics Letters. 2015 Jul 27;107(4):041101.

[4] Zektzer R, Talker E, Barash Y, Mazurski N, Stern L, Levy U. Atom–Photon Interactions in Atomic Cladded Waveguides: Bridging Atomic and Telecom Technologies. ACS Photonics. 2021 Feb 26;8(3):879-86.

[5] Ritter R, Gruhler N, Pernice WH, Kübler H, Pfau T, Löw R. Coupling thermal atomic vapor to an integrated ring resonator. New Journal of Physics. 2016 Oct 20;18(10):103031.

[6] ] Naiman A, Sebbag Y, Levy U. Demonstration of on Chip Cavity to Thermal Rb Vapor Strong Coupling. In2020 Conference on Lasers and Electro-Optics (CLEO) 2020 May 10 (pp. 1-2). IEEE.

[7] H. Raether, Surface Plasmons, Springer-Verlag, Berlin, 1988.

[8] Stern L, Grajower M, Levy U. Fano resonances and all-optical switching in a resonantly coupled plasmonic–atomic system. Nature communications. 2014 Sep 8;5(1):1-9.

[9] MOSLEH M, HAMIDI S, RANJBARAN M. Multifunctional Logic Gates based on Resonant Transmission at Atomic-Plasmonic structure.

[10] Kuttge M, Vesseur EJ, Verhoeven J, Lezec HJ, Atwater HA, Polman A. Loss mechanisms of surface plasmon polaritons on gold probed by cathodoluminescence imaging spectroscopy. Applied Physics Letters. 2008 Sep 15;93(11):113110.

[11] Yeh P, Yariv A, Cho AY. Optical surface waves in periodic layered media. Applied Physics Letters. 1978 Jan 15;32(2):104-5.

[12] R. Dubey, E. Barakat, M. Häyrinen, M. Roussey, S. Honkanen, M. Kuittinen and H. P. Herzig, “Experimental investigation of the propagation properties of Bloch surface waves on dielectric multilayer platform,” JEOS:RP 13(1), 5 (2017).

[13] Villa F, Regalado LE, Ramos-Mendieta F, Gaspar-Armenta J, Lopez-Ríos T. Photonic crystal sensor based on surface waves for thin-film characterization. Optics letters. 2002 Apr 15;27(8):646-8

[14]Shinn M, Robertson WM. Surface plasmon-like sensor based on surface electromagnetic waves in a photonic band-gap material. Sensors and Actuators B: Chemical. 2005 Mar 28;105(2):360-4.

[15] Konopsky VN, Alieva EV. Photonic crystal surface waves for optical biosensors. Analytical chemistry. 2007 Jun 15;79(12):4729-35.

[16] Michelotti F, Sciacca B, Dominici L, Quaglio M, Descrovi E, Giorgis F, Geobaldo F. Fast optical vapour sensing by Bloch surface waves on porous silicon membranes. Physical Chemistry Chemical Physics. 2010;12(2):502-6.

[17] Liscidini M, Galli M, Shi M, Dacarro G, Patrini M, Bajoni D, Sipe JE. Strong modification of light emission from a dye monolayer via Bloch surface waves. Optics letters. 2009 Aug 1;34(15):2318-20.

[18] Pirotta S, Xu XG, Delfan A, Mysore S, Maiti S, Dacarro G, Patrini M, Galli M, Guizzetti G, Bajoni D, Sipe JE. Surface-enhanced Raman scattering in purely dielectric structures via Bloch surface waves. The Journal of Physical Chemistry C. 2013 Apr 4;117(13):6821-5.

[19] Delfan A, Liscidini M, Sipe JE. Surface enhanced Raman scattering in the presence of multilayer dielectric structures. JOSA B. 2012 Aug 1;29(8):1863-74.

[20] Angelini AN, Enrico E, De Leo N, Munzert P, Boarino L, Michelotti F, Giorgis F, Descrovi E. Fluorescence diffraction assisted by Bloch surface waves on a one-dimensional photonic crystal. New Journal of Physics. 2013 Jul 2;15(7):073002.

[21] Frascella F, Ricciardi S, Pasquardini L, Potrich C, Angelini A, Chiadò A, Pederzolli C, De Leo N, Rivolo P, Pirri CF, Descrovi E. Enhanced fluorescence detection of miRNA-16 on a photonic crystal. Analyst. 2015;140(16):5459-63.

[22] Xiang Y, Tang X, Fu Y, Lu F, Kuai Y, Min C, Chen J, Wang P, Lakowicz JR, Yuan X, Zhang D. Trapping metallic particles using focused Bloch surface waves. Nanoscale. 2020;12(3):1688-96.

[23] Yu L, Barakat E, Sfez T, Hvozdara L, Di Francesco J, Peter Herzig H. Manipulating Bloch surface waves in 2D: a platform concept-based flat lens. Light: Science & Applications. 2014 Jan;3(1):e124-.

[24] Dubey R, Lahijani BV, Barakat E, Häyrinen M, Roussey M, Kuittinen M, Herzig HP. Near-field characterization of a Bloch-surface-wave-based 2D disk resonator. Optics letters. 2016 Nov 1;41(21):4867-70.

[25] Menotti M, Liscidini M. Optical resonators based on Bloch surface waves. JOSA B. 2015 Mar 1;32(3):431-8.

[26] Dubey R, Lahijani BV, Häyrinen M, Roussey M, Kuittinen M, Herzig HP. Ultra-thin Bloch-surface-wave-based reflector at telecommunication wavelength. Photonics Research. 2017 Oct 1;5(5):494-9.

[27] Gulkin DN, Popkova AA, Afinogenov BI, Shilkin DA, Kuršelis K, Chichkov BN, Bessonov VO, Fedyanin AA. Mie-driven directional nanocoupler for Bloch surface wave photonic platform. Nanophotonics. 2021 Sep 1;10(11):2939-47.

[28] Descrovi E, Sfez T, Quaglio M, Brunazzo D, Dominici L, Michelotti F, Herzig HP, Martin OJ, Giorgis F. Guided Bloch surface waves on ultrathin polymeric ridges. Nano letters. 2010 Jun 9;10(6):2087-91.

[29] Wu X, Barakat E, Yu L, Sun L, Wang J, Tan Q, Herzig HP. Phase-sensitive near field Investigation of Bloch surface wave propagation in curved waveguides. Journal of the European Optical Society-Rapid publications. 2014 Oct 29;9.

[30] Stehle C, Bender H, Zimmermann C, Kern D, Fleischer M, Slama S. Plasmonically tailored micropotentials for ultracold atoms. Nature Photonics. 2011 Aug;5(8):494-8.

[31]Stehle C, Zimmermann C, Slama S. Cooperative coupling of ultracold atoms and surface plasmons. Nature Physics. 2014 Dec;10(12):937-42.

[32] Ritter R, Gruhler N, Pernice WH, Kübler H, Pfau T, Löw R. Coupling thermal atomic vapor to an integrated ring resonator. New Journal of Physics. 2016 Oct 20;18(10):103031.

[33] Naiman A, Sebbag Y, Talker E, Barash Y, Stern L, Levy U. Large cooperativity in strongly coupled chip-scale photonic-atomic integrated system.

[34] Gryga M, Vala D, Kolejak P, Gembalova L, Ciprian D, Hlubina P. One-dimensional photonic crystal for Bloch surface waves and radiation modes-based sensing. Optical Materials Express. 2019 Oct 1;9(10):4009-22.

[35] Limonov MF, Rybin MV, Poddubny AN, Kivshar YS. Fano resonances in photonics. Nature Photonics. 2017 Sep;11(9):543-54.

[36] Scheel S, Buhmann SY, Clausen C, Schneeweiss P. Directional spontaneous emission and lateral Casimir-Polder force on an atom close to a nanofiber. Physical Review A. 2015 Oct 15;92(4):043819.